\documentclass[]{spie}  

\usepackage{amsmath,amsfonts,amssymb}
\usepackage{graphicx}
\usepackage{aas_macros}
\usepackage[colorlinks=true, allcolors=blue]{hyperref}

\title{Status of the Planet Formation Imager (PFI) concept \footnote{Copyright 2016 Society of Photo-Optical Instrumentation Engineers. One print or electronic copy may be made for personal use only. Systematic reproduction and distribution, duplication of any material in this paper for a fee or for commercial purposes, or modification of the content of the paper are prohibited. DOI: http://dx.doi.org/10.1117/12.2233926}}

\author[a]{Michael J. Ireland}
\author[b]{John D. Monnier}
\author[c]{Stefan Kraus}
\author[d]{Andrea Isella}
\author[e]{Stefano Minardi}
\author[f]{Romain Petrov}
\author[g]{Theo ten Brummelaar}
\author[h]{John Young}
\author[i]{Gautum Vasisht}    
\author[j]{David Mozurkewich} 
\author[k]{Stephen Rinehart}  
\author[l]{Ernest A. Michael}
\author[m]{Gerard van Belle}
\author[n]{Julien Woillez}
\affil[a]{Research School of Astronomy and Astrophysics, Australian National University, Canberra, ACT 2611, Australia}
\affil[b]{Department of Astronomy, University of Michigan, Ann Arbor, Michigan 48109, USA}
\affil[c]{School of Physics, University of Exeter, Exeter, EX4 4QL, UK}
\affil[d]{Rice University, Department of Physics and Astronomy, Houston, TX, 77521}
\affil[e]{Leibniz-Institut f\"{u}r Astrophysik Potsdam (AIP), Germany}
\affil[f]{Universit\'{e} C\^{o}te d'Azur/OCA/CNRS, Parc Valrose, F-06108 Nice, France}
\affil[g]{CHARA Array, Georgia State University, USA}
\affil[h]{Cavendish Laboratory, University of Cambridge, J J Thompson Avenue, Cambridge, CB3 0HE, UK}
\affil[i]{Jet Propulsion Laboratory, California Institute of Technology, 4800 Oak Grove Drive, Pasadena, CA 91109}
\affil[j]{Seabrook Engineering, 9310 Dubarry Ave, Seabrook, MD 20706, USA}
\affil[k]{NASA Goddard Space Flight Center, Mail Code 665, Greenbelt, MD, 20771}
\affil[l]{Dept. of Electrical Engineering, FCFM, University of Chile, Av. Tupper 2007, Santiago, Chile}
\affil[m]{Lowell Observatory, 1400 W. Mars Hill Road, Flagstaff, AZ, USA}
\affil[n]{European Southern Observatory (ESO), Karl-Schwarzschild-Str. 2, Garching 85748, Germany}

\authorinfo{Further author information: (Send correspondence to M.J.I.)\\M.J.I.: E-mail: michael.ireland@anu.edu.au, Telephone: +61 2 612 50288}

\begin{document} 
\maketitle

\begin{abstract}
The Planet Formation Imager (PFI) project aims to image the period of planet assembly directly, resolving structures as small as a giant planet's Hill sphere. These images will be required in order to determine the key mechanisms for planet formation at the time when processes of grain growth, protoplanet assembly, magnetic fields, disk/planet dynamical interactions and complex radiative transfer all interact -- making some planetary systems habitable and others inhospitable. We will present the overall vision for the PFI concept, focusing on the key technologies and requirements that are needed to achieve the science goals. Based on these key requirements, we will define a cost envelope range for the design and highlight where the largest uncertainties lie at this conceptual stage. 
\end{abstract}


\section{INTRODUCTION}
\label{sec:intro}  

The Planet Formation Imager (PFI) is a science-driven concept first formally presented at the 2014 SPIE conference\cite{Monnier14,Kraus14,Ireland14}. The most basic premise of the concept is that planet formation is a process so complex that understanding the processes at work in individual systems and the solar system's cosmic context requires imaging. The key spatial scale is the {\em Hill Sphere} radius, which is the radius of gravitational sphere of influence of a forming planet. This is 2.5\,milli-arcsec for a Jupiter analog in the nearest star forming regions at $\sim$140\,pc, or 0.5\,milli-arcsec for a Jupiter-like planet at 1\,AU separations. 

In this paper, we will present the driving science requirements in Section~\ref{sec:requirements}; the technology downselect in Section~\ref{sec:TechDownselect}, focusing on the signal-to-noise of the different options; a brief end-to-end description of a baseline design in Section~\ref{sec:baseline};  a largely top-down cost model in Section~\ref{sec:costing}; and conclusions and future work in Section~\ref{sec:conclusions}. There are several companion papers in these proceedings, including Kraus et al.\cite{kra16} (Scientific Context), Monnier et al.\cite{mon16} (Technical Roadmap), Minardi et al. (Beam Combination), Petrov et al. (Fringe Tracking), Vasisht et al. (Heterodyne technology), Baron et al. (Imaging Algorithms) and Mozurkewich et al. (Beam Transport).

\section{DRIVING SCIENCE REQUIREMENTS}
\label{sec:requirements}

PFI must be able to study planet formation from birth through to mature planetary systems, at ages of $<$1\,Myr through to $>$100\,Myr. The draft driving science requirements, as well as current goals, which may become requirements, are given in Table~\ref{tab:my_label}.

\begin{table}
\centering
\begin{tabular}{|c|c|}
\hline
 Parameter & Range \\
\hline
Wavelengths & 8-13\,$\mu$m \\
Source Distance & 140\,pc \\
Spatial Resolution & 0.7\,mas \\
Spectral Resolving Power & R$>$20 \\
FOV & $>$0.15" \\
Sensitivity (integrated flux) & 100\,mJy@11\,$\mu$m (N$\sim$7) \\
Star magnitude & $m_H\sim$9 \\
Maximum star angular size & $>$0.15\,mas \\
\hline
 Goal Wavelengths & 3-30\,$\mu$m \\
 Goal Source Distances & 50--500\,pc \\
 Goal Spatial Resolution & 0.2\,mas \\
 Goal Spectral Resolution & R$\sim$100,000 \\
 Goal Surface brightness (K) & 150\,K \\
 Goal L' 5-$\sigma$ sensitivity (mag) & 18.5 \\
\hline
\end{tabular}
\caption{Key parameters for PFI}
\label{tab:my_label}
\end{table}

The driving scientific need for high angular resolution is resolving circumplanetary emission from features of e.g.~planetary winds or circumstellar disk features due to planetary formation. The key length scale is the {\rm Hill sphere} radius (e.g.~see Monnier et al. 2016 and Kraus et al. 2016 in these proceedings). One driving reason for observations in the $N$ bandpass (8--13\,$\mu$m) is the desire to image even embedded planet formation, where potentially even the circumplanetary disk is extincted, and all that is seen is a warm region of the protoplanetary disk. The size of such a warm region is of order the disk scale height, which is similar in scale to the Hill sphere radius for Jovian-mass planets. 

The driving scientific need for high spectral resolution (listed as a goal in Table~\ref{tab:my_label}) is the ability to kinematically resolve circumplanetary emission from other disk emission, as well as measuring the masses of many young planets through spectro-astrometry or resolved disk emission. A characteristic minimum spectral resolving power for circumplanetary disk emission is 30,000, corresponding to 10\,km/s and the orbital velocity of Jupiter's most massive moon, Ganymede. 

Sensitivity requirements are further driven by two other key science requirements: detection of a ``typical'' point-source exoplanet across a range of ages, and detecting the passively heated structure of a protoplanetary disk at least out to the ``ice-line'' at 150\,K. The chosen goal limit of L'=18.5 corresponds to a 1\,M$_J$ COND\cite{Baraffe03} model of 100\,Myr at 17\,pc (the distance to several already known AB~Dor moving group members), a 1\,M$_J$ COND model of 10\,Myr at 64\,pc (slightly more distant than TW~Hya), and an accreting protoplanetary disk with $M \dot{M} \sim$1--3 $\times 10^{-6}$\,M$_\odot^2$/yr at the distance to Taurus or $\rho$~Ophiucus (depending on detailed assumptions) \cite{Zhu15}.

In addition to these requirements, there is a goal for observations of key prototype objects including HL~Tau, TW~Hya and LkCa~15. LkCa~15 is well represented by these requirements, while HL~Tau requires either infrared fringe tracking or a laser guide star, and TW~Hya has a slightly larger stellar angular size of 0.19\,mas and a brighter $m_H$ of 7.6, potentially affecting fringe tracking requirements.

\section{TECHNOLOGY DOWNSELECT}
\label{sec:TechDownselect}

The ultimate location for PFI is in space, where spacecraft such as Spitzer and Herschel have shown that instrumental backgrounds approaching the zodiacal background are possible. However, kilometric-scale formation flying of more than 2 telescopes in an array pointed in an arbitrary direction presents a very significant challenge of technological readiness. The most ambitious space-based interferometer launched so far has been the LISA pathfinder, at $\sim$US\,\$460 million, which included only the formation flying and metrology aspects of PFI, for only 2 telescopes. The infrared detection capabilities of PFI are arguably encapsulated by a mission on the scale of the Spitzer space telescope cost, which was $\sim$US\,\$720 million. Concept studies for a less ambitious non-formation flying interferometer such as SPIRIT have spoken about a ``probe'' class mission within NASA. Taken together, this top-down approach to PFI indicates that the best approach at the present time is likely to be an initial focus on a significantly less expensive ground-based concept, and continued engagement by the PFI technical and scientific working groups at a low level with teams proposing space-based interferometer concepts.

For 9--13\,$\mu$m science, two concepts are possible: direct-detection and heterodyne interferometry. For less than $\sim$20 telescopes, direct detection is advantageous under almost any assumptions. For shorter wavelengths (e.g.~the CO fundamental band at 4.7\,$\mu$m), direct detection offers a clear advantage, and heterodyne detection is uncompetitive. We will therefore focus first on the direct detection design for PFI, then consider where heterodyne technologies may augment such a concept.

\subsection{DIRECT DETECTION SIGNAL-TO-NOISE}

Mid-infrared signal to noise considerations for a heterodyne architecture were presented in Ireland and Monnier (2014)\cite{Ireland14}. In this section, we describe point-source detection signal-to-noise, focusing on the case of background-limited observations. 

\subsubsection{SNR Derivation}

Direct detection SNR for a variety of interferometric beam combiners can be derived by starting with the photometric SNR for a single telescope, expressed simply as the number of target photons per unit time and fractional bandwidth divided by the square root of the shot noise variance sum of the target and the background:

\begin{align}
( \frac{S}{N} )_p &= \frac{S(\lambda) 10^{-0.4m} A \eta_w \eta_c}{\sqrt{S(\lambda) 10^{-0.4m} A \eta_w \eta_c + B(T,\lambda)\eta_c (1-\eta_w)}} \sqrt{t} \sqrt{\frac{\Delta\lambda}{\lambda}},
\label{eqnSNRPhotometric}
\end{align}

where $A$ is the telescope area, $\eta_w$ is the warm instrument efficiency, $\eta_c$ is the cold instrument efficiency, $t$ the integration time, $\Delta \lambda$ the bandwidth, $\lambda$ the wavelength, $m$ the apparent (Vega) magnitude and $N_t$ the number of telescopes. $S(\lambda)$ is the photon rate per unit fractional bandwidth per unit area per unit time of Vega \cite{Tokunaga05,Hewett06,Cohen92}, and $B(T,\lambda)$ is the blackbody function expressed in terms of photons per unit fractional bandwidth per (dual polarization) spatial mode:

\begin{align}
B(T,\lambda) &= \frac{2 c}{\lambda}\frac{1}{\exp(hc/\lambda k_B T) - 1}
\end{align}

This formula assumes single-mode beam combination, with coupling loss most likely part of the warm optics emissivity $\eta_w$ (although this is dependent on the detailed design). For a single baseline, the signal and noise is split equally (in time or space) between measuring the real and imaginary components of the visibility. Irrespective of the beam combiner details, any idea combiner (neglecting uncertainties in measuring total flux) has the point source visibility signal-to-noise:

\begin{align}
( \frac{S}{N} )_d &= ( \frac{S}{N} )_p~~~{\rm (single~baseline)}
\label{eqnSingleBaseline}
\end{align}

For more than one baseline, the SNR depends on the combiner architecture. 
For an all-in-one combiner, the signal per baseline is independent of $N_t$, and the noise
(due to the target and background) is proportional to $\sqrt{N_t}$. For a pairwise combiner,
each telescope's signal is split $N_t-1$ ways, reducing per-baseline SNR by $\sqrt{N_t-1}$. Irrespective of the combiner type,the point-source detection signal to noise increases as $\sqrt{N_b}=\sqrt{N_t(N_t-1)/2}$, with $N_b$ the number of baselines. This gives:

\begin{align}
( \frac{S}{N} )_d &= \sqrt{N_t-1} ( \frac{S}{N} )_p~~~{\rm (all-in-one)} \\
( \frac{S}{N} )_d &= \sqrt{N_t/2}( \frac{S}{N} )_p~~~{\rm (pairwise)} 
\label{eqnMultipleBaseline}
\end{align}

The fundamental reason for this difference is that in our analysis, the all-in-one combiner was assumed not to have photometric taps. Such a scheme would rely on {\em self-calibration} for imaging fidelity, which loses little imaging signal-to-noise for large $N_t$. If half the light were to be used for photometric taps, the two schemes would have equal signal-to-noise in the limit of large numbers of baselines. Note that if target shot noise dominates, higher SNR can be achieved when searching for faint companions by using nulling architectures.

Another beam combining possibility is a direct imaging, densified pupil design. The key limitation of a direct imaging instrument is signal-to-noise in the case of a sparse (e.g. non-redundant) pupil. The fraction of the light in the central core is proportional to $(N_{\rm tel} D^2 / B_{\rm max}^2) \times (\gamma_D/\gamma_B)^2$, where $D$ is the telescope diameter, $B_{\rm max}$ the maximum baseline and $\gamma_D$ and $\gamma_B$ the magnification factors for telescope pupils and baselines respectively. For example, for a 9-telescope non-redundant array\cite{Golay71}, this fraction is 0.36, or 0.326 once the geometric factor of circular pupils fitting in a hexagonal grid is taken into account. In general, $(D/B_{\rm max}) \times (\gamma_D/\gamma_B) < \sqrt{\pi/N(N-1)} $. For background-limited direct imaging or imaging faint structures around a bright star, the signal-to-noise is reduced by this factor. Recovering this signal-to-noise is possible by analysing the direct image using methods analogous to aperture-mask interferometry, however those techniques require $\lambda/\Delta\lambda > (B_{\rm max}/D) \times (\gamma_B/\gamma_D)$. For the fields of view required for PFI, this either means another significant sensitivity loss with small $\Delta \lambda$, or an integral field unit, removing the simplicity of the direct imaging scheme and making the beam combiner similar to any other all-in-one combiner.

Although a densified, redundant pupil does have high signal-to-noise, the number of telescopes required is $\pi (B_{\rm max}/D)^2$, which is more than 300 for the PFI designs considered here. This is impractical given the need for fringe tracking.

\begin{align}
( \frac{S}{N} )_d &= N_t (\frac{S}{N} )_p~~~{\rm (fully~densified~pupil)} \\
( \frac{S}{N} )_d &\le \pi (\frac{S}{N} )_p~~~{\rm (densified~non-redundant~pupil,~}N_t \ge 6)
\label{eqnDensified}
\end{align}

\subsubsection{SNR For Key Filters}
\label{sec:SNR}

Considering the highest signal-to-noise design for a non-redundant pupil, the {\em all-in-one} design, and considering a scenario limited by thermal background, the point source detection Signal-to Noise Ratio (SNR) is:

\begin{align}
( \frac{S}{N} )_d &= C(T,\lambda) 10^{-0.4m} A  \frac{\eta_w}{\sqrt{1-\eta_w}}\sqrt{\eta_c} \sqrt{t} \sqrt{\frac{\Delta \lambda}{\lambda}} \sqrt{N_t - 1},
\label{eqnSNRDirect}
\end{align}

where $A$ is the telescope area, $\eta_w$ is the warm instrument efficiency, $\eta_c$ is the cold instrument efficiency, $t$ the integration time, $\Delta \lambda$ the bandwidth, $\lambda$ the wavelength, $m$ the apparent (Vega) magnitude and $N_t$ the number of telescopes. $C(T,\lambda)$ is a function of warm optics temperature and central wavelength $\lambda$, given in Table~\ref{tabC}. The background temperature and filter combinations are noted in the table where the background-limited assumption breaks down. In those cases, a nulling beam-combiner design would be needed in order to maintain full sensitivity.

Table~\ref{tabMagLim} shows the magnitude limits based on this analysis for two different array size assumptions.

\begin{table}
\caption{The function $C(T,\lambda)$ ($m^{-2}s^{-1/2}$) in Equation~\ref{eqnSNRDirect} for the L', M and N bandpasses. Atmospheric efficiency was included and was based on 10\,mm of precipitable water vapour.}
\begin{tabular}{llllll}
\hline
Bandpass & Wavelength & Bandwidth &  T=280\,K & T=270\,K & T=220\,K \\
& ($\mu$m)  & ($\mu$m) & & & \\
\hline
L' & 3.7  & 0.8 & 2.42e+05 & 3.14e+05 & 1.61e+06 \\
M & 5.0  & 0.8 & 1.45e+04 & 1.76e+04 & 5.91e+04 \\
N & 10.5 & 5.0 & 6.78e+02 & 7.43e+02 & 1.33e+03 \\
\hline
\end{tabular}
\label{tabC}
\end{table}

\begin{table}
\caption{Limiting apparent magnitudes for 5-$\sigma$ point-source detections for two PFI array sizes\textbf{n} based on the the bandpasses in Table~\ref{tabC}, assuming $\eta_w=0.39$, $\eta_c=0.44$, and 10$^4$\,s integration. The telescope diameter and number of telescopes assumed is given in the first column. }
\begin{tabular}{lllllll}
\hline
Assumptions & Bandpass & Wavelength & Bandwidth &  T=280\,K & T=270\,K & T=220\,K \\
& & ($\mu$m)  & ($\mu$m) & & & \\
\hline
2.5\,m, N$_t$=12 & L' & 3.7 & 0.8 &  17.7 &  18.0 &  19.8$^b$ \\
& M & 5.0  & 0.8 &  14.5 &  14.7 &  16.0 \\
& N & 10.5 & 5.0 &  11.8 &  11.9 &  12.5 \\
4\,m, N$_t$=21  & L' & 3.7 & 0.8 &  19.1$^b$ &  19.3$^b$ &  21.1$^a$ \\
& M & 5.0  & 0.8 &  15.8 &  16.0 &  17.4 \\
& N & 10.5 & 5.0 &  13.1 &  13.2 &  13.8 \\
\hline
\end{tabular}
\newline
$^a$: Requires nulling of starlight for essentially all PFI targets (solar-type stars in nearby star forming regions)\\
$^b$: Requires nulling of starlight for brighter PFI targets to achieve this sensitivity.
\label{tabMagLim}
\end{table}

\begin{table}
\caption{3-$\sigma$ surface brightness sensitivity of PFI, all for an assumed warm optics temperature of 270\,K. }
\begin{tabular}{lllllll}
\hline
Assumptions & Bandpass & Wavelength & Bandwidth & B$_{\rm max}$=1\,km & B$_{\rm max}$=2\,km & B$_{\rm max}$=4\,km \\
 & & ($\mu$m)  & ($\mu$m) & & & \\
\hline
2.5\,m, N$_t$=12 & L' & 3.7 & 0.8 &  289 &   323 &   364 \\
 & M & 5.0  & 0.8 &  268 &   308 &   361 \\
 & N & 10.5 & 5.0 &  154 &   183 &   224 \\
4\,m, N$_t$=21  & L' & 3.7 & 0.8 &  265 &   293 &   327 \\
 & M & 5.0  & 0.8 &  240 &   272 &   313 \\
 & N & 10.5 & 5.0 &  135 &   157 &   187 \\
\hline
\end{tabular}
\label{tabSurfaceBrightness}
\end{table}

\begin{table}
\caption{Top of atmosphere photon rate per unit area per unit fractional bandwidth for Vega, $S(\lambda)$ \cite{Tokunaga05,Hewett06,Cohen92}. }
\begin{tabular}{lllllll}
\hline
Filter 				 & H    & K & L' & M' & N & Q \\
Wavelength ($\mu$m)  & 1.65 & 2.2 & 3.7 & 4.8 & 10.5 & 21 \\
$S$					& 1.5e+10 & 9.5e+09 & 3.7e+09 & 2.3e+09 & 5.3e+08 & 1.3e+08 \\
\hline 
\end{tabular}
\label{tabSTable}
\end{table}

\section{BASELINE PFI DESIGN}
\label{sec:baseline}

\subsection{ARRAY ARCHITECTURE}
\label{sec:architecture}

Array architecture considerations are detailed in Monnier et al. in these proceedings. In brief, the maximum stellar angular size ($>$0.15\,mas) combined with a fringe tracking wavelength of 1.65\,$\mu$m, a minimum fringe-tracking visibility of 0.5 and the need for non-catastrophic failure if one telescope has a drop-out or is offline, means that the maximum separation of second nearest-neighbor telescopes is 1.6\,km. For large arrays, this requirement naturally lends itself to ring or Y configurations. For maximum baselines less than 3.2\,km, a wider range of array geometries are possible.

\subsection{SITE SELECTION}
\label{sec:site}

The scientific considerations of site selection are detailed in Kraus et al. in these proceedings. A mid-latitude site is strongly preferred, as this enables access to both the Taurus and $\rho$~Ophiucus star forming regions. Nearby young moving groups are also mostly located in the southern hemisphere, meaning that a southern site is slightly preferred over a northern site. This being said, it is possible a mid-latitude northern site, a mid-latitude southern site or even an Antarctic site could provide a sufficient number of targets to answer the key scientific goals of PFI.

\begin{figure}
\includegraphics[width=0.48\textwidth]{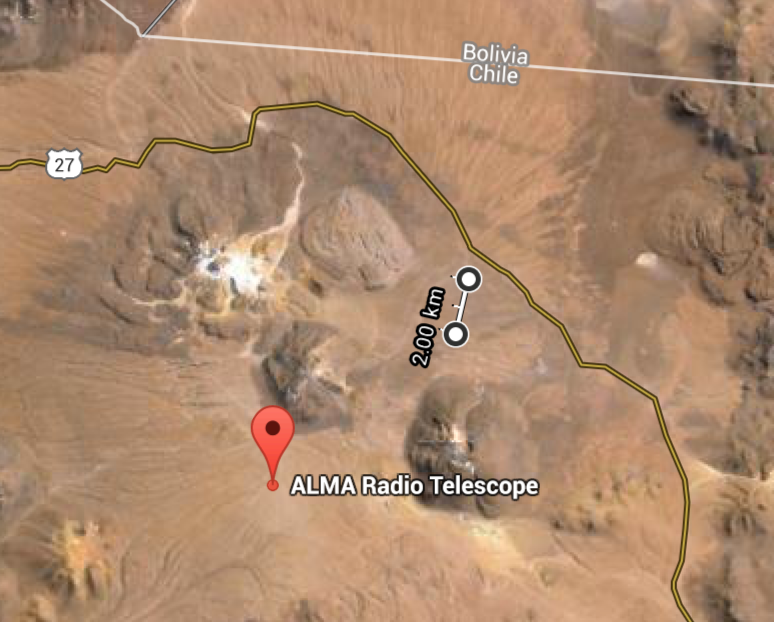}
\includegraphics[width=0.48\textwidth]{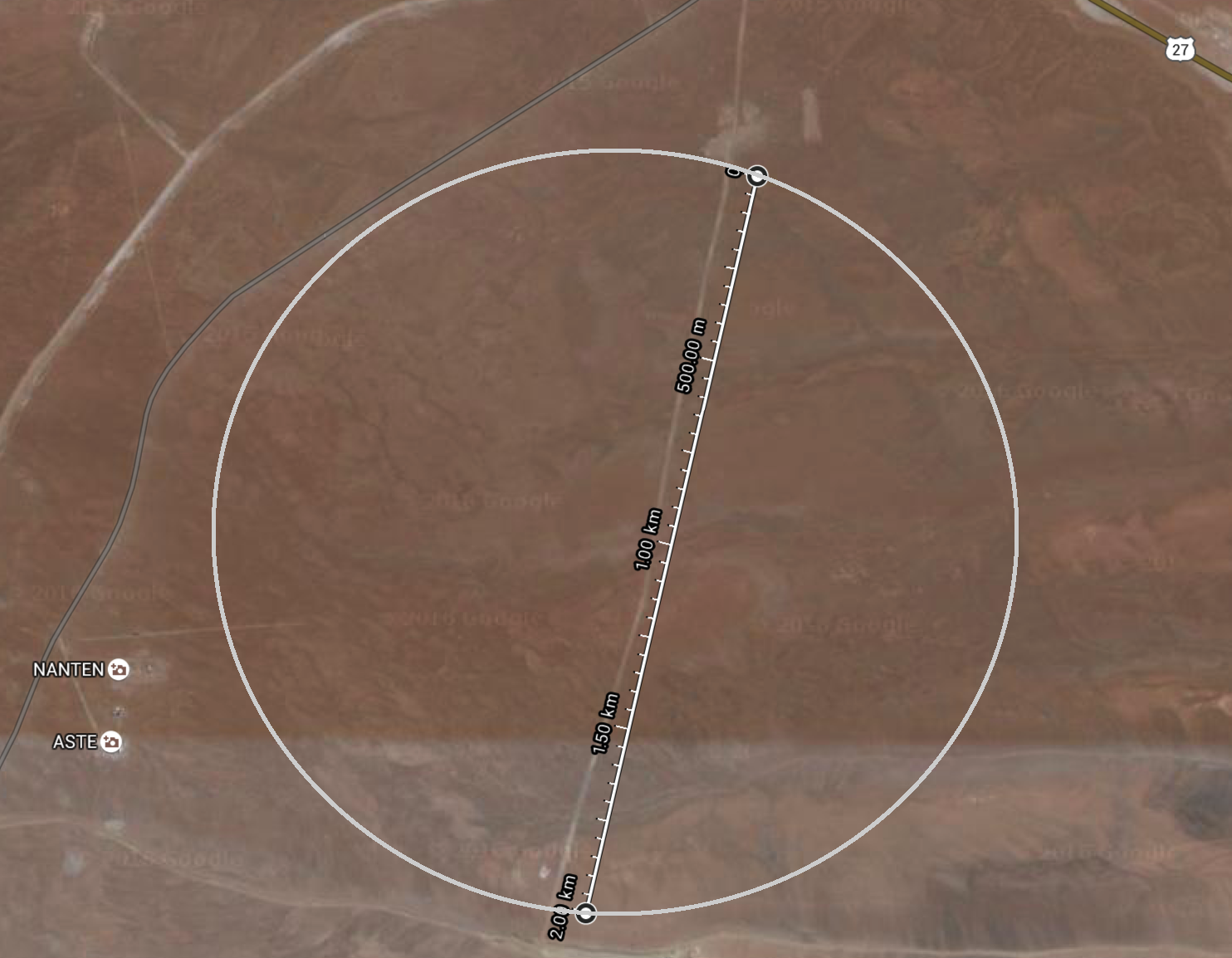}
\caption{An example suitable site for PFI, next to the ALMA observatory. This site has low water vapour (essential for $\sim$5\,$\mu$m observations) and sub-arcsecond seeing expected. With a circular array geometry, pylon heights would be less than 10\,m (credit: Google maps web services).}
\label{figChajnantor}
\end{figure}

For the most important astronomical filters, H and K for fringe-tracking, and L' and N for science, Precipitable Water Vapour (PWV) is not a critical site requirement, and any site with $<$10\,mm, $\sim$80\% of the time is adequate. However, for the infrared M band (4.6-5.4\,$\mu$m) and the infrared Q band (24.1-24.8\,$\mu$m), PWV becomes a critical requirement, and a site with $<$1\,mm PWV is strongly preferred. The seeing parameters $r_0$ and $t_0$ are both reasonably important, as fringe-tracking sensitivity is proportional to H and K-band Strehl ratios. For a direct detection array in particular, a reasonably flat site is needed. A site near the ALMA at Chajnantor, for example that shown in Figure~\ref{figChajnantor}, meets these requirements.

\subsection{TELESCOPES}
\label{sectTelescopes}

We will consider a nominal telescope diameter of 2.5\,m, which is likely to be adequate for fringe tracking on HL~Tau (depending on inner disk versus central star emission), and certainly adequate for fringe tracking on solar-type stars without an inner disk such as LkCa~15. Given the complexities of fringe tracking, including tracking in the presence of adaptive optics drop-outs and realistically uneven transmission of telescopes, a detailed simulation is needed to really determine the true magnitude limit as a function of diameter.

\begin{figure}
\includegraphics[width=0.8\textwidth]{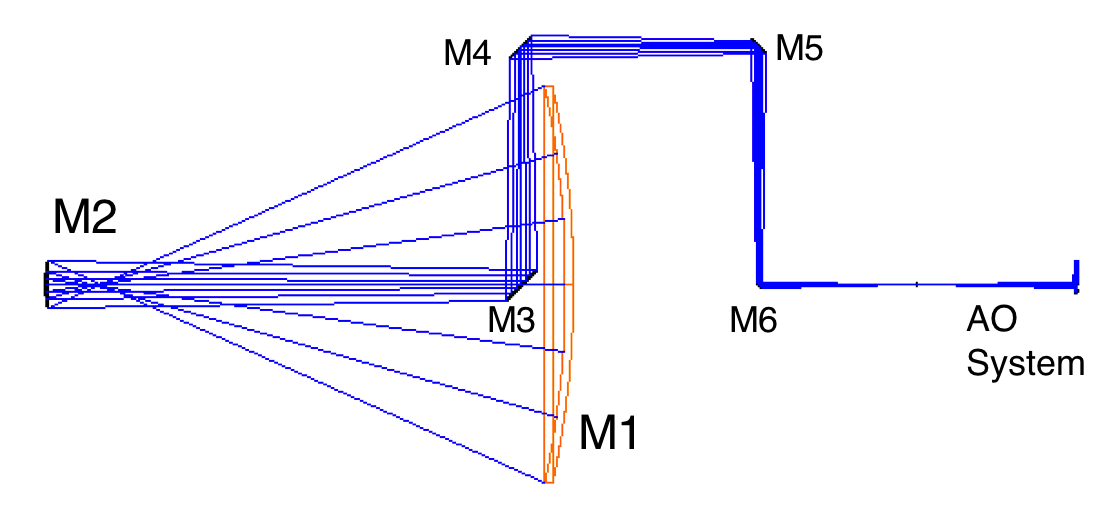}
\caption{Side-on optical design of a spherical primary Gregorian telescope with a ``conventional'' beam relay for an altitude-azimuth design. This is the baseline architecture considered here -- an altitude-altitude design and/or deformable secondary may improve performance (i.e. fewer reflections) but may also add cost.}
\label{figTelZemax}
\end{figure}

This telescope size is at the range where there are a great number of altitude-azimuth professional telescopes around the world, and at the upper end of ``standard products'' for major telescope manufacturers. A nearly conventional optical design is shown in Figure~\ref{figTelZemax}, with the beam relay following the elevation (M3 and M4) and azimuth (M6 and M7) axes. With a fast primary mirror focal ratio, this design typically requires a very stiff structure to hold the secondary mirror, which can be large for moderate field of view telescopes. PFI is potentially significantly simpler, with a small field of view that could be served well by a small secondary mirror and a spherical primary mirror. As the primary mirror in such a design has no optical axis, only the secondary tilt has to be controlled to compensate for any lateral translation of the secondary. This significantly reduces requirements for mount stiffness. Another variant on this design is a Gregorian design, where the pupil becomes naturally apodized, reducing diffractive losses and enabling higher fiber coupling efficiencies.

Finally, the lack of Nasmyth instruments and the fast focal ratios enabled by the small field of view mean that a relatively small dome could be adequate for the PFI telescopes. The expected dome to telescope diameter ratio would then be smaller than that of Keck (3.7), resulting in anticipated domes at the high end of  near ``standard products'' from major manufacturers, at the $<$US\,\$150 $\times 10^3$ level. For these reasons, in the cost envelope section~\ref{sec:costing}, both conventional and 50\% cost telescopes are considered.

\subsection{BEAM TRANSPORT}

Beam transport for a direct detection PFI concept is certainly challenging. The beam area required for any given interferometric architecture is proportional to $B\lambda$ for a baseline $B$ and wavelength $\lambda$. This means that by this metric, PFI is $\sim$10 times more challenging than VLTI or $\sim$35 times more challenging than CHARA. In these proceedings, Mozurkewich et al. in these proceedings discuss potentials for beam transport, including the potential for both beam transport to the beam combining laboratory and the delay line to exist in a single pipe. This is the design we consider as baseline here, with all wavelengths travelling along the same beam train, as it requires no substantially new technological elements. Achieving a smooth motion of the delay cart is certainly challenging, especially with pipe length changes due to diurnal temperature fluctuations needing to be accommodated, and likely pipe irregularity requiring larger amplitude pupil offset control than the MROI delay lines. Other concepts discussed in Mozurkewich et al. in these proceedings may be less expensive and lower risk, but do require additional optical surfaces and in some cases the complexity of either very broad antireflection coatings or multiple delay lines for different wavelengths.

\begin{figure}
\includegraphics[width=\textwidth]{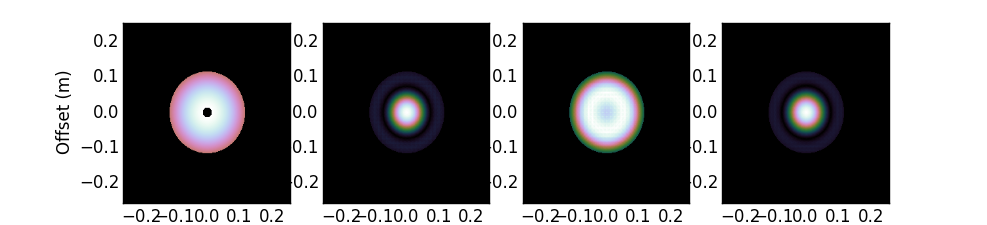}
\includegraphics[width=\textwidth]{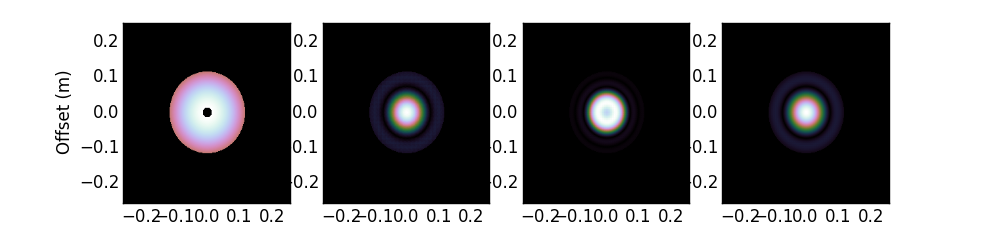}
\includegraphics[width=\textwidth]{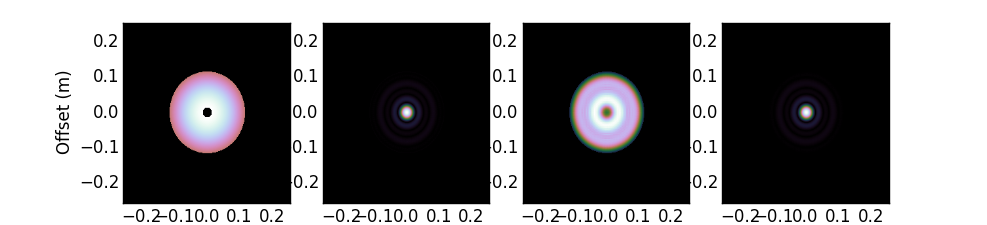}
\caption{Beam intensities at the telescope, the fixed catseye, the moving catseye and the beam reducing confocal paraboloid pair, for a configuration with a beam diameter of 0.23\,m and: $\lambda$=13\,$\mu$m, $R$=1\,km and $d$=950\,m (top row, 90\% transmission);  $\lambda$=13\,$\mu$m, $R$=1\,km and $d$=550\,m (middle row, 93\% transmission); 
$\lambda$=5\,$\mu$m, $R$=1\,km and $d$=950\,m (bottom row, 96\% transmission). The apodized telescope pupil is due to the assumption of a fast Gregorian telescope with spherical primary (Section~\ref{sectTelescopes}).}
\label{figBeamTransportDiffraction}
\end{figure}

Figure~\ref{figBeamTransportDiffraction} shows a simulation of diffraction through this beam transport arrangement. With a beam diameter of only 230\,mm, diffractive losses are less than 10\% for all delay line positions at 13\,$\mu$m, so long as the pupil is apodized, due to e.g. a Gregorian telescope design with a spherical primary. Technological prototyping for such a design would be necessary for power transport, speed and beam height servoeing, but there are no showstoppers identified, and standard 0.6\,m vacuum rated pipe could be used.

%

\subsection{FRINGE TRACKING}

A variety of different fringe tracking architectures are discussed in Petrov et al. in these proceedings. For the baseline PFI concept considered here, it is possible that the central star is only partially resolved at even the longest baseline. In this situation, for a photon-limited fringe tracker, there is relatively little difference in notional sensitivity for a variety of designs. Where all baselines are used, signal-to-noise per baseline is significantly reduced, but (as in the Gravity combiner), if information from multiple baselines are optimally combined and readout noise is negligible, the signal-to-noise per delay line is recovered.

Irrespective of these details, at least one fringe tracking concept offers N-telescope fringe tracking signal-to-noise that is equal to 2-telescope fringe-tracking signal-to-noise. This concept is the hierarchical fringe tracker, which has as few as $N_{\rm tel}+1$ outputs prior to dispersion. A photonic version of this combiner would be a tri-coupler, where the two outer guides of the input are populated by nominally in-phase inputs, resulting in most of the input flux leaving the combiner in the central guide of the coupler output. Fringe tracking simply consists of balancing the flux in the two outer guides of the coupler output.

\subsection{SCIENCE COMBINERS}


In these proceedings, Minardi et al. illustrate and describe a series of beam combination 
schemes and available technologies for mid-infrared multi-telescope combination. 
Simulations show that, for identical throughput, pairwise ABCD\cite{Benisty:2009} or 
DBC\cite{Minardi:2010} combination schemes are more sensitive for individual baseline visibility amplitudes than multi-axial, all-in-one combiners, with details of this comparison depending on the utility of photometric taps in the final image reconstruction. 
Given the presence of adaptive optics correction at each PFI telescope, bulk optics combiners could offer higher throughput respect to photonic ones, as modal filtering may not be strictly necessary for high precision visibility measurements.
A bulk optics pairwise combiner would almost certainly be prohibitively complex for $\ge$12 telescopes, as appears to be needed for PFI. An all-in-one combiner, on the other hand, adds no additional optics per beam when increasing the number of beams. One of the simplest examples of an all-in-one combiner is an architecture where a linear non-redundant array \cite{Bedding94,Ireland08} is dispersed in an orthogonal direction. These designs necessarily require cylindrical optics, with an anamorphic factor of order $N_{\rm baselines}$ needed. 

In terms of costing and performance, the key metrics are throughput and cost. A bulk optics design could be made with a grating cross disperser and $\sim$12 optical surfaces, with a total cold efficiency budget of 0.44, including detector quantum efficiency. The volume is dominated by the cross-disperser size (of order 100\,mm for a high-dispersion combiner at $\sim$5\,$\mu$m), and by the need to close-pack $N_{\rm tel}$ beams without diffraction becoming a problem. For 12 telescopes, at the most challenging wavelength of 13\,$\mu$m, required pupil sizes are also only 100mm. These are smaller than typical numbers for 8\,m class telescope instruments, and do not represent significant new technological challenges.
Given that a bulk optics design requires no new technology, it can be seen as a baseline for 3 combiners at L', M and N bands (covering 3--13\,$\mu$m). 

However, the same SNR performance could be expected also for a photonic combiner, so long as a comparable efficiency was obtained. Further potential advantages of the photonic platform are the possibility for considerable volume reduction (influencing cost) without throughput loss, higher precision where visibilities are dominated by a bright central point source, 
and nulling capabilities at shorter wavelengths where background noise may not dominate.
A current drawback of this approach is the low maturity of integrated optics at mid-infrared wavelengths, as compared to the telecom bands. However, recent progress in the field (see Labadie et al. in these proceedings for a review) shows that this solution could be as well become competitive within the time horizon of the PFI project.

\section{COST ENVELOPE}
\label{sec:costing}

\subsection{Direct Detection and Common Elements}

At such an early stage, there is significant uncertainty in the cost model for PFI. This section aims to show some justifiable numbers, and how they scale with telescope diameter, number of telescopes and ``big-picture'' specifications for PFI. This enables a quantitative technology downselect, and a rough optimal cost to place on efficiency. The first key element of the cost model is scaling the cost of opto-mechanical assemblies (e.g. telescopes) for multiple units. For small numbers, the once-off design cost is saved, and for larger numbers, there are significant savings with e.g. custom fixtures and tooling for optical and mechanical assembly, integration and test. This assumed functional form is:

\begin{align}
f_{\rm mult}(N_{\rm tel}) &= 0.3 N_{\rm tel} + 0.7  (1+\log_e(N_{\rm tel}))
\end{align}

This model is loosely based on experience with large instrument projects, as well as the break between e.g. the Planewave instruments SDK700, and the formula for single larger telescopes applied to a nearly ``once-off'' 0.7\,m design \cite{Swift14}. Our telescope cost model for ``standard'' telescopes is:

\begin{align}
C_{T, {\rm standard}} &= {\rm US\,}\$1.3 \times 10^6  \times f_{\rm mult}(N_{\rm tel}) D^2.
\end{align}

We will assume that $f_{\rm mult}$ applies to the optics and moving components of beam transport, but not the large mechanical components which are dominated by standard large engineering installation of vacuum pipes. For diffraction reasons, the diameter of beam transport pipes are proportional to $\sqrt{B_{\rm max}}$, meaning that the material cost per unit length for hardware is proportional to $B_{\rm max}$ and the overall cost proportional to $B_{\rm max}^2$. For the pipes itself, we assume labour costs equal to material costs of pipes and supports. We also assume that there is a fixed (\$1\,USM) design and prototyping phase, with the delay cart and metrology cost being linearly proportional to baseline length. As a rough guess, the cart itself is \$50K of mechanical parts, up to 15 actuators at \$5K per actuator, \$60K in optics, \$50K in metrology, \$10K in computing and electronics and double this in assembly, integration and test effort. We then arrive at:

\begin{align}
C_{D} &= {\rm US\,}\$0.8 \times 10^6 N_{\rm tel} \times (\frac{B_{\rm max}}{2\,{\rm km}})^2 + {\rm US\,}\$0.5 \times 10^6 \times f_{\rm mult}(N_{\rm tel}) \times (0,5 + 0.5\frac{B_{\rm max}}{2\,{\rm km}}) + {\rm US\,}\$1 \times 10^6.
\end{align}

Based on the recent experience of the CHARA array, we model the adaptive optics cost as:

\begin{align}
C_{AO} &= (0.5 + 0.1D^2) \times f_{\rm mult}(N_{\rm tel}) \times {\rm US\,}\$1\times 10^6.
\end{align}

The beam combiner cost is difficult to estimate, especially given the wide range of costs for {\em state of the art} beam combiners at CHARA and VLTI. Assuming a baseline cost compatible with a facility but relatively simple single-mode 8m telescope instrument of \$6M for each of the fringe tracker, H/K and N band detectors, we model the beam combiner cost as:

\begin{align}
C_{BC} &= (12 + 0.3N_{\rm tel}) \times {\rm US\,}\$1\times 10^6.
\end{align}

The minimum cost for the beam combiner central building is of order \$400K for 200\,m$^2$ of floor space, plus a small increase for additional beams. Note that the delay architecture is assumed to be inside the beam transport pipes. We then arrive at only:

\begin{align}
C_{\rm Building} &= (0.4 + 0.02N_{\rm tel}) \times {\rm US\,}\$1\times 10^6.
\end{align}

Note that in practice, existing interferometers add office buildings and sometimes accommodation to this cost. Using remote operation and an existing site would eliminate much of this additional cost in buildings.

The cost of access roads for an interferometer could vary by a large factor, depending in detail on the nature of the site. We will assume a cost of \$0.5M per km for the site main access road, and \$0.1M per km for telescope access roads (i.e.~one lane gravel). This results in:

\begin{align}
C_{\rm Infrastructure} &= (0.1N_{\rm tel} (\frac{B_{\rm max}}{1\,km}) + 0.5  (\frac{B_{\rm max}}{1\,km})) \times {\rm US\,}\$1\times 10^6.
\end{align}

Finally, the total project cost should include an additional 10\% management overhead and 20\% contingency. 

\subsection{Heterodyne Detection}

A heterodyne design saves on beam transport costs and direct detection costs, but adds significant costs at each telescope with a high dispersion spectrograph and linear detector array. Other approaches could include using integrated optics and mid-IR fiber technology, as well as a potential increase in the bandpass of each spectral channel with new detector technologies. As many of these technologies do not yet exist, some of these costs are even less certain than the direct detection costs. Some of these new technologies are discussed in Vasisht et al. in these proceedings. The beam transport cost is significantly smaller in the Heterodyne design, and we ignore the baseline-dependent costs:

\begin{align}
C_{D, {\rm het}} &= {\rm US\,}\$0.2 \times 10^6 \times N_{\rm tel}  + {\rm US\,}\$0.3 \times 10^6 \times f_{\rm mult}(N_{\rm tel})  + {\rm US\,}\$1\times 10^6.
\end{align}

The heterodyne disperser at each telescope is similar to a small instrument on a large telescope. We assume a cost for this element that (optimistically) assumes a very small cost component for the heterodyne mixer:

\begin{align}
C_{\rm spect} &= {\rm US\,}\$2 \times 10^6 \times f_{\rm mult}(N_{\rm tel}) 
\end{align}

The computing cost for the heterodyne array scales as the number of baselines, with a 30 telescope array considered in Ireland and Monnier (2014) \cite{Ireland14}. We assume:

\begin{align}
C_{\rm computing} &= {\rm US\,}\$10 \times 10^6 \times (\frac{N_{\rm tel}}{30})^2
\end{align}

Finally, the beam combining cost is roughly reduced by a factor of 3 in the purely heterodyne concept, because there is no L' and M or N-band combiner, but  there a H and/or K band fringe tracker is still required. This gives:

\begin{align}
C_{\rm BC, {\rm het}} &= (4 + 0.1\times N_{\rm tel}) \times {\rm US\,}\$1\times 10^6.
\end{align}

Like the direct detection and common element costs, we add 10\% management overhead and 20\% contingency to these costs.

\subsection{Cost Comparison}

\begin{figure}
\includegraphics[width=0.44\textwidth]{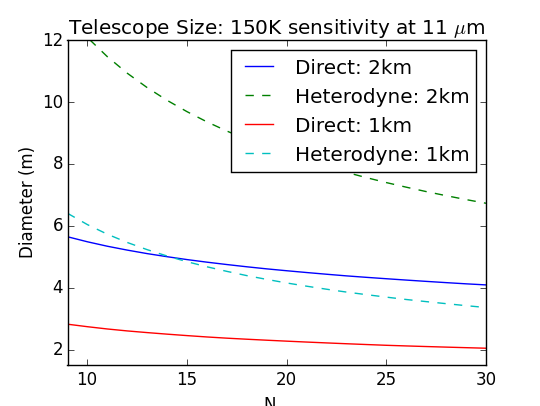}
\includegraphics[width=0.55\textwidth]{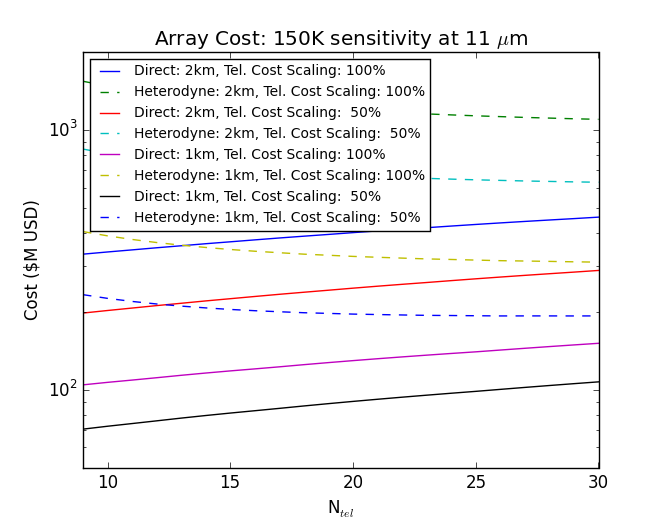}
\caption{Left: Required telescope diameter to achieve a 3-$\sigma$ 150\,K surface brightness sensitivity in the 8-13\,$\mu$m band, for different maximum baselines and a 10$^4$\,s integration time. Assumed array efficiencies are: warm efficiency $\eta_w=0.39$, cold efficiency $\eta_c=0.44$, heterodyne efficiency $\eta_{\rm het}=0.35$, heterodyne bandpass coverage fraction 50\%. Coherent integration from short-wavelength tracking is assumed (i.e. with no dependence on seeing for infrared sensitivity). Right: Cost model results for these arrays. 
Although heterodyne does not appear competitive in this bandpass, it is much more competitive in the 24.5\,$\mu$m window. The dominance of the telescope costs can be seen in the models where the telescope costs are reduced by 50\%, which may be possible with new, narrow field of view, lightweight designs.}
\label{fig:150K}
\end{figure}

\begin{figure}
\includegraphics[width=0.44\textwidth]{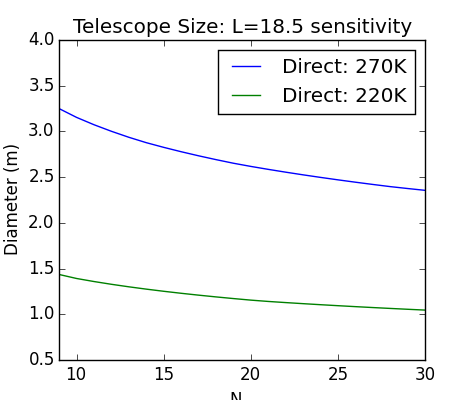}
\includegraphics[width=0.55\textwidth]{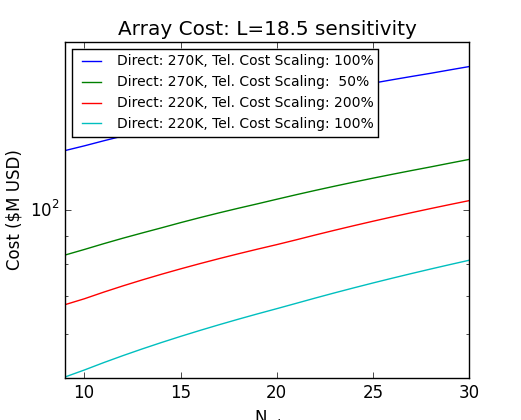}
\caption{Left: Required telescope diameter to achieve a 5-$\sigma$ L'=18.5 point-source detection limit in a 10$^4$\,s integration time, for warm optics temperatures corresponding to Chajnator (270\,K) and the high antarctic plateau (220\,K). Other parameters as for Figure~\ref{fig:150K}. Right: Cost model results for these arrays. Even if telescopes are assumed to be twice as expensive in Antarctica and half as expensive for new technology telescopes in a temperate site, Antarctica is still a more cost-effective site for the L' bandpass.}
\label{fig:L185}
\end{figure}

Armed with these (mostly top-down) guesses of subsystem costs, we can examine the different cost of architectures that meet the science requirements. Figures~\ref{fig:150K} and~\ref{fig:L185} show telescope diameter and total costs for PFI to meet the surface brightness and point-source detection sensitivity requirements, using the equations in \cite{Ireland14}, Section~\ref{sec:SNR} and Section~\ref{sec:costing}. Surface brightness was converted to flux using the Planck formula and an assumption of a circular beam with diameter $\lambda/B_{\rm max}$. These figures demonstrate the following conclusions:

\begin{itemize}
\item At 11\,$\mu$m, for $N_{\rm tel}<30$ it is clear that heterodyne is not competitive.
\item Overall array costs drives the direct detection design to the smallest number of moderately large telescopes.
\item Large maximum baselines are expensive to maintain surface brightness sensitivity. As long as maximum baselines are little more than 1\,km, the L' point-source detection limit drives costs.
\item Telescopes drive the overall cost of PFI, and reducing the costs by a factor of 2 over a ``conventional'' design (as may be possible) reduces PFI costs by $\sim$33\%.
\item PFI should be achievable for a total price tag of less than US\,\$150 million.
\end{itemize}

Finally, note that heterodyne is much more competitive than direct detection for $N_t \ge 9$ in the Q band (24.1-24.8\,$\mu$m), and easily achieves the 150\,K sensitivity, all be it at a lower spatial resolution. As this technology is far from mature, it hasn't been considered in the cost model.

\subsection{Optimal Price for Efficiency}

Given this {\em big-picture} cost model, we can ask questions such as: ``For a cost-optimal design at fixed signal-to-noise ratio, what is the incremental cost placed on an increase in warm optics throughput $\eta_w$?''. 

For the baseline design with 12 telescopes of 2.5\,m diameter, we note that the cost is roughly proportional to the product of telescope area and warm efficiency $\eta_w$. In this case, the optimal price is placed on improving efficiency if the marginal costs for a relative increase in area and warm efficiency are equal:

\begin{align}
\frac{dC}{d \log{A}} &= \frac{dC}{d \log{\eta_w}}
\end{align}

This results in perhaps surprisingly large numbers, with US\,\$60k per percent of warm efficiency per telescope for warm optics, and US\,\$260k per percent of efficiency for cold beam-combiner optics (for either all beam-combiners combined, or the one with the scientifically most critical signal to noise). Armed with this knowledge, trade-offs become significantly easier in considering changes to a baseline design. For example, removing 4 optics in the telescope beam train for an MROI-like design is worthwhile only if it costs significantly less than \$720\,K per telescope (assuming a 12\% efficiency hit from those reflections). Similarly, changing coatings to aluminium to enable auxiliary science also has a cost associated with it.

\section{CONCLUSIONS}
\label{sec:conclusions}

The Planet Formation Imager (PFI) baseline design as presented here consists of $\sim$12 telescopes of $\sim$2.5\,m diameter in a non-redundant configuration (possibly a circular or Y-shaped array), with 0.9--1.3\,$\mu$m adaptive optics sensing wavelengths, 1.5--2.4$\mu$m fringe tracking wavelengths, and 3--13\,$\mu$m direct detection, all-in-one combiner science wavelengths. At this stage, for an array of this scale, heterodyne beam combination is unlikely to be a competitive technology at 8--13\,$\mu$m, but more work is certainly needed in this area. Heterodyne beam combination at 24.5\,$\mu$m is extremely competitive, but that technology does not yet exist -- any PFI design should certainly include room at telescopes for future heterodyne detectors.

Such an interferometer would be able to detect young forming Jupiter-mass exoplanets in the nearest star forming regions, and follow their evolution over a period of $>$100\,Myr. In addition, it would be able to image protoplanetary disk structures right down to the temperature of the ice line at 150\,K, and down to angular scales corresponding to the {\em Hill sphere}, the gravitational sphere of influence of a planet. Finally, in favourable situations, PFI will be able to determine the masses of forming exoplanets from velocity-resolved observations of the circumplanetary disks. These observations will be essential to complete our emerging picture of planet formation, currently derived from indirect observations.


\acknowledgments 
 
 The authors thank the broader technical working group and science working group teams in developing the technical and scientific working group draft white-papers, on which part of this was based.
 

\bibliography{report} 
\bibliographystyle{spiebib} 

\end{document}